\documentclass[seceq]{ptptex}

\title{A Variational Principle for Dissipative Fluid Dynamics}

\author{Hiroki \textsc{Fukagawa}%
\thanks{E-mail: hiroki@beer.appi.keio.ac.jp%
} and Youhei \textsc{Fujitani}%
\thanks{E-mail: youhei@appi.keio.ac.jp%
} }

\inst{School of Fundamental Science \& Technology, Keio University,\\
Yokohama 223-8522, Japan}

\abst{
In the variational principle leading to the Euler equation for a perfect
fluid, we can use the method of undetermined multiplier for holonomic
constraints representing mass conservation and adiabatic condition.
For a dissipative fluid, the latter condition is replaced by the
constraint specifying how to dissipate. Noting that this constraint
is nonholonomic, we can derive the balance equation of momentum for
viscous and viscoelastic fluids by using a single variational
principle. We can also derive the associated Hamiltonian formulation
by regarding the velocity field as the input in the framework of control
theory.
}
\begin{document} 
\maketitle

\section{Introduction\label{sec:Introduction}}

In many areas of physics, variational principles help us choose convenient variables in formulating problems. 
The least-action principle  can lead to the equations of motion not
only for microscopic dynamics but also those for nondissipative macroscopic
dynamics. For example, in a perfect fluid, the sum of the kinetic
energy and internal energy is conserved, and we can write down
the action easily. There are also the conservation laws for mass and
entropy; these holonomic constraints can be incorporated into the
action by means of the method of undetermined multipliers. Thus, the
Euler equation can be derived from the stationary condition of the
action\cite{Bateman,Lin,SELINGER,Bennett:,FukagawaFujitani,Schutz,Salmon,Kambe,Yoshida}.

The Navier-Stokes equation is the equation of motion for the Newtonian
viscous fluid, and represents the momentum balance with the dissipative
force taken into account in the linear phenomenological law\cite{onsager,Degroot}.
Without this force, the equation is reduced to the Euler equation.
The dissipative force can be derived using another variational
principle called the maximum dissipation principle, where Rayleigh's
dissipation function is minimized\cite{Goldstein,Serrin}. Using this
principle amounts to assuming the linear phenomenological law. 
A set of variables with respect to which Rayleigh's
dissipation function is minimized is usually different from a set of variables with respect to which
the action is stationary \cite{nicolis,Martyusheva}.

One way of unifying these two variational principles is called Onsager's
variational principle \cite{doibook}, which has been used frequently
in formulating dissipative dynamics of complex fluids \cite{doi90,DoiOhnuki,doi,liusensei,liu2}.
Using this principle, we can obtain the balance equation of momentum
with the dissipative force taken into account in the linear phenomenological
law. However, applying this principle to the Newtonian fluid in a
straightforward way, we can obtain only a linearized version of Navier-Stokes
equation, i.e., the Stokes equation\cite{Takaki}.

A viscous fluid is not locally adiabatic, and the entropy conservation
law cannot be assumed. We instead impose a constraint specifying how to dissipate. Noting that this constraint is nonholonomic, unlike
in the previous studies, we can derive the balance equation of momentum
in the framework of a single variational principle. As mentioned above,
this framework for a viscous fluid originates from that for a perfect fluid, which we
briefly review in \S \ref{sub:Variational-perfect}. Our main results
are shown in \S \ref{sub:Dissipative}, where we apply our procedure
to viscous and viscoelastic fluids. In \S \ref{sec:Hamiltonian-formulations-for},
for completeness, we derive the associated Hamiltonian formulation
in the framework of the control theory. Our study is discussed and
summarized in the last section.

\section{Perfect fluid\label{sub:Variational-perfect}}

We briefly review Ref.~\citen{FukagawaFujitani} with slight
modifications. Let us consider the dynamics of a perfect fluid in a
fixed container from the initial time $t_{{\rm init}}$ to the final
time $t_{{\rm fin}}$. Let $V$ denote the region occupied by the
fluid. The velocity field $\boldsymbol{v}$ satisfies no-penetration
condition at the boundary $\partial V$, i.e., 
\begin{equation}
\boldsymbol{n}\cdot\boldsymbol{v}=0\ {\rm on}\ \partial V\ ,\label{eq:boundary of v}
\end{equation}
 where $\boldsymbol{n}$ is the unit normal vector directed outside
on $\partial V$. Let us write $\rho$ for mass per unit volume, and
$s$ for entropy per unit \emph{mass}. The internal-energy density
per unit \emph{mass}, $\epsilon$, is a function of $\rho$ and $s$
because of the local equilibrium. The action is given by the integral
of the Lagrangian density with respect to the time and space considered.
Apart from the constraints mentioned below, the Lagrangian density
is given by 
\begin{equation}
{\cal L}(\rho,\boldsymbol{v},s)\equiv\rho\left\{ \frac{1}{2}\boldsymbol{v}^{2}-\epsilon(\rho,s)\right\} \ ,\label{eq:intro}
\end{equation}
 which is the difference between kinetic energy density and internal
energy density. The variables are not independent because of the constraints
mentioned below. Writing $T$ and $p$ for the temperature and the
pressure, respectively, we have $d\epsilon=-pd\rho^{-1}+Tds$ in the
thermodynamics, which leads to 
\begin{equation}
p\equiv\rho^{2}\left(\frac{\partial\epsilon}{\partial\rho}\right)_{s}\ {\rm and}\ T\equiv\left(\frac{\partial\epsilon}{\partial s}\right)_{\rho}\ ,\label{eq:Temp}
\end{equation}
 where the subscripts $_{s}$ and $_{\rho}$ indicate variables fixed
in the respective partial differentiations. We write $\tau$ for the
time in the Lagrangian coordinates although it is equivalent to the
time $t$ in the non-relativistic theory. The partial derivatives
with respect to $\tau$~($\partial_{\tau}$) and $t$~($\partial_{t}$)
imply the Lagrangian and Eulerian time-derivatives, respectively.

In the Lagrangian description, we label a fluid particle with its
initial position $\boldsymbol{a}=(a_{1},a_{2},a_{3})$, and write
$\boldsymbol{X}=(X_{1},X_{2},X_{3})$ for its position at time $\tau$.
The time derivative of $\boldsymbol{X}$ denotes the velocity fields
$\boldsymbol{v}$, i.e., 
\begin{equation}
\partial_{\tau}\boldsymbol{X=\boldsymbol{v}}\ .\label{eq:X}
\end{equation}
 The endpoints of the path line are fixed by 
\begin{equation}
\delta\boldsymbol{X}(\boldsymbol{a},t_{{\rm init}})=\delta\boldsymbol{X}(\boldsymbol{a},t_{{\rm fin}})=\boldsymbol{0}\ .\label{eq:boundaries for X}
\end{equation}
 The volume element in the Lagrangian coordinates can be given by
the determinant of the Jacobian matrix, 
\begin{equation}
J(\boldsymbol{a},\tau)\equiv\frac{\partial(X_{1},X_{2},X_{3})}{\partial(a_{1},a_{2},a_{3})}\ .\label{eq:jacobian}
\end{equation}
 We assume that $J$ has no singular points in the space and time
considered and that $J(\boldsymbol{a},t_{{\rm init}})$ is unity.
The conservation law of mass is given by 
\begin{equation}
\rho(\boldsymbol{a},\tau)J(\boldsymbol{a},\tau)-\rho_{{\rm init}}(\boldsymbol{a})=0\ ,\label{eq:mass}
\end{equation}
 where $\rho_{{\rm init}}(\boldsymbol{a})$ is the initial value of
mass density. The adiabatic condition implies that the entropy of
a fluid particle in the perfect fluid is conserved along a path line,
\begin{equation}
s(\boldsymbol{a},\tau)-s_{{\rm init}}(\boldsymbol{a})=0\ ,\label{eq:adiabatic}
\end{equation}
 where $s_{{\rm init}}(\boldsymbol{a})$ is the initial value of entropy
density per mass. Equations \eqref{eq:mass} and \eqref{eq:adiabatic}
give holonomic constraints. Using undetermined multipliers,
$\boldsymbol{\gamma}$, $K$ and $\Lambda$, we can define the action
as 
\begin{equation}
\int_{t_{{\rm init}}}^{t_{{\rm fin}}}\!\!\!\!\!\! d\tau\!\int_{V}\!\! d^{3}\boldsymbol{a}\ \left\{ J{\cal L}(\rho,s,\boldsymbol{v})+\!\boldsymbol{\gamma}\cdot(\partial_{\tau}\boldsymbol{X}-\boldsymbol{v})+\! K(\rho J-\rho_{{\rm init}})\!+\!\Lambda J(s-s_{{\rm init}})\right\} \ ,\label{eq:L-action}
\end{equation}
 which is a functional of $\boldsymbol{\gamma}$, $K$, $\Lambda$,
$\rho$, $s$, $\boldsymbol{v}$ and $\boldsymbol{X}$. Taking the
variations of Eq.~\eqref{eq:L-action} with respect to these trajectories,
respectively, leads to Eqs.~\eqref{eq:X}, \eqref{eq:mass}, \eqref{eq:adiabatic},
\begin{eqnarray}
K & = & -\frac{1}{2}\boldsymbol{v}^{2}+\epsilon+\frac{p}{\rho}\ ,\label{eqn:rho}\\
\Lambda & = & \rho T\ ,\label{eq:T}\\
\boldsymbol{\gamma} & = & \rho J\boldsymbol{v}\label{eqn:vel}
\end{eqnarray}
 and 
\begin{equation}
\frac{\partial}{\partial\tau}\gamma_{i}=-\frac{\partial}{\partial a_{j}}\left\{ \rho\left(\frac{1}{2}\boldsymbol{v}^{2}-\epsilon+K\right)\frac{\partial J}{\partial(\partial X_{i}/\partial a_{j})}\right\} \ .\label{eqn:varilag}
\end{equation}
 Roman indices run from 1 to 3, and repeated indices are summed up.
The surface integral terms, appearing on the way of this calculation,
vanish because of the boundary conditions, Eqs.~\eqref{eq:boundary of v}
and \eqref{eq:boundaries for X}. Calculating by means of the cofactors
yields 
\begin{equation}
\frac{\partial J}{\partial(\partial X_{i}/\partial a_{j})}=J\frac{\partial a_{j}}{\partial X_{i}}\ ,
\end{equation}
 while some algebra yields 
\begin{equation}
\frac{\partial}{\partial a_{j}}\left(J\frac{\partial a_{j}}{\partial X_{1}}\right)=0\ .
\end{equation}
 Thus, we can rewrite Eq.~\eqref{eqn:varilag} as
\begin{equation}
\frac{\partial}{\partial\tau}\gamma_{i}=-J\frac{\partial}{\partial X_{i}}\left\{ \rho\left(\frac{1}{2}\boldsymbol{v}^{2}-\epsilon+K\right)\right\} \ .\label{eqn:varilag-1}
\end{equation}
 Substituting Eqs.~\eqref{eqn:rho} and \eqref{eqn:vel} into Eq.~\eqref{eqn:varilag-1},
we successfully obtain the Euler equation in the Lagrangian description,
\begin{equation}
\rho\frac{\partial v_{i}}{\partial\tau}=-\frac{\partial p}{\partial X_{i}}\ .\label{eqn:Eulereq}
\end{equation}
 We can replace $\partial v_{i}/\partial\tau$ by $(\partial t+\boldsymbol{v}\cdot\nabla)v_{i}$
to obtain the Euler equation in the Eulerian description, which is
given by Eq.~\eqref{eq:E-Euler} below\cite{SELINGER,Bennett:}.

In the Eulerian description, a fluid particle is labeled by the Lagrangian
coordinates $\boldsymbol{A}\equiv(A_{1},A_{2},A_{3})$, which depend
on the spatial position $\boldsymbol{x}$. Since the Lagrangian coordinates
are conserved along the path line, we have 
\begin{equation}
\frac{\partial}{\partial t}A_{i}=-\boldsymbol{v}\cdot\nabla A_{i}\ .\label{eq:fixing endpoints}
\end{equation}
 The endpoints of a path line are fixed by 
\begin{equation}
\delta\boldsymbol{A}(\boldsymbol{x},t_{{\rm init}})=\delta\boldsymbol{A}(\boldsymbol{x},t_{{\rm fin}})=\boldsymbol{0}\ ,\label{eq:boundaries for endpoints in the Eulerian}
\end{equation}
 instead of Eq.~\eqref{eq:boundaries for X}, as discussed in Ref.~\citen{FukagawaFujitani}.
The mass conservation law Eq.~\eqref{eq:mass} can be rewritten as
\begin{equation}
\rho(\boldsymbol{x},t)=J^{-1}(\boldsymbol{x},t)\rho_{{\rm init}}(\boldsymbol{A}(\boldsymbol{x},t))\ ,\label{eq:mascon}
\end{equation}
 where the inverse of $J$ is given by 
\begin{equation}
J^{-1}(\boldsymbol{x},t)=\frac{\partial(A_{1},A_{2},A_{3})}{\partial(x_{1},x_{2},x_{3})}\ .\label{eq:invJ}
\end{equation}
 The action in the Lagrangian description, Eq.~\eqref{eq:L-action},
is rewritten in the Eulerian description as
\begin{equation}
\int_{t_{{\rm init}}}^{t_{{\rm fin}}}\!\!\!\! dt\int_{V}\!\!\! d^{3}\!\boldsymbol{x}\ \left\{ {\cal L}(\rho,s,\boldsymbol{v})+\!\beta_{i}(\partial_{t}A_{i}+\boldsymbol{v}\cdot\nabla A_{i})+K(\rho-\rho_{{\rm init}}J^{-1})+\Lambda(s-s_{{\rm init}})\right\} ,\label{eq:E-action}
\end{equation}
 which is a functional of $\boldsymbol{\beta}$, $K$, $\Lambda$,
$\rho$, $s$, $\boldsymbol{v}$ and $\boldsymbol{A}$. The stationary
condition of Eq.~\eqref{eq:E-action} yields Eqs.~\eqref{eq:adiabatic},
\eqref{eqn:rho}, \eqref{eq:T}, \eqref{eq:fixing endpoints}, \eqref{eq:mascon},
\begin{equation}
\rho\boldsymbol{v}+\beta_{i}\nabla A_{i}=\boldsymbol{0}\label{eq:v}
\end{equation}
 and 
\begin{equation}
\frac{\partial}{\partial t}\beta_{i}=-\nabla\cdot(\beta_{i}\boldsymbol{v})-KJ^{-1}\frac{\partial\rho_{{\rm init}}}{\partial A_{i}}+\frac{\partial}{\partial x_{j}}\left\{ K\rho_{{\rm init}}\frac{\partial J^{-1}}{\partial(\partial A_{i}/\partial x_{j})}\right\} -\Lambda\frac{\partial s_{{\rm init}}}{\partial A_{i}}\ .\label{eq:betaa}
\end{equation}
 Here, $\rho_{{\rm init}}$ and $s_{{\rm init}}$ depend on $\boldsymbol{A}$,
but $\rho$ and $s$ are independent variables. Using Eqs.~\eqref{eq:adiabatic},
\eqref{eqn:rho}, \eqref{eq:T}, 
\begin{equation}
\frac{\partial J^{-1}}{\partial(\partial A_{i}/\partial x_{j})}=J^{-1}\frac{\partial x_{j}}{\partial A_{i}}
\end{equation}
 and 
\begin{equation}
\frac{\partial}{\partial x_{j}}\left(J^{-1}\frac{\partial x_{j}}{\partial A_{i}}\right)=0\ ,
\end{equation}
 we can rewrite Eq.~\eqref{eq:betaa} as 
\begin{eqnarray}
\frac{\partial}{\partial t}\beta_{i} & = & -\nabla\cdot(\beta_{i}\boldsymbol{v})+\left(\rho\frac{\partial K}{\partial x_{j}}-\rho T\frac{\partial s}{\partial x_{j}}\right)\frac{\partial x_{j}}{\partial A_{i}}\nonumber \\
 & = & -\nabla\cdot(\beta_{i}\boldsymbol{v})+\left(-\frac{\rho}{2}\frac{\partial\boldsymbol{v}^{2}}{\partial x_{j}}+\frac{\partial p}{\partial x_{j}}\right)\frac{\partial x_{j}}{\partial A_{i}}\ .\label{eq:beta-1}
\end{eqnarray}
 Dividing Eq.~\eqref{eq:v} by $\rho$, we calculate the Lie derivative
of the resulting equation with respect to $\boldsymbol{v}$ to find
the sum of the Lie derivative of $\boldsymbol{v}$ and 
\begin{equation}
\left(\nabla A_{i}\right)\left(\frac{\partial}{\partial t}+\boldsymbol{v}\cdot\nabla\right)\left(\frac{\beta_{i}}{\rho}\right)\label{eqn:lie}
\end{equation}
 to vanish, with the aid of Eq.~\eqref{eq:fixing endpoints}. The
Lie derivative of $\boldsymbol{v}$ was given in Appendix A of
Ref.~\citen{FukagawaFujitani}. Substituting Eq.~\eqref{eq:beta-1}
into Eq.~\eqref{eqn:lie}, we can derive the Euler equation, 
\begin{equation}
\rho\left\{ \frac{\partial}{\partial t}\boldsymbol{v}+\frac{1}{2}\nabla\boldsymbol{v}^{2}-\boldsymbol{v}\times(\nabla\times\boldsymbol{v})\right\} =-\nabla p\ .\label{eq:E-Euler}
\end{equation}

\section{Dissipative fluids\label{sub:Dissipative}}

\subsection{Simple dissipative system\label{simple}}

To show the essence of our procedure, we consider a damped harmonic
oscillator in a heat bath. We define the position of the oscillator,
$q$, so that it vanishes in the balance. We write $m$ and $k$ for
the mass of the oscillator and the spring constant, respectively.
The internal energy of the heat bath, $E$, is a function of its entropy,
$S$, and its temperature is given by $T\equiv dE/dS$. We assume
that the frictional force, $f$, is exerted on the oscillator. The
action is given by the integral of the difference between the kinetic
energy and the sum of the potential energy and the internal energy
over the time interval $[t_{{\rm init}},t_{{\rm fin}}]$,

\begin{equation}
\int_{t_{{\rm init}}}^{t_{{\rm fin}}}\!\!\!\! dt\left\{ \frac{1}{2}m\left(\frac{dq}{dt}\right)^{2}-\left(\frac{1}{2}kq^{2}+E(S)\right)\right\} \ ,\label{eqn:Oaction}
\end{equation}
 which is a functional of trajectories $q(t)$ and $S(t)$. The frictional
force converts the mechanical energy into heat. Thus, the realized
trajectories should satisfy 
\begin{equation}
f\frac{dq}{dt}+T\frac{dS}{dt}=0\ ,\label{eqn:friction}
\end{equation}
 considering that the heat bath is always at the equilibrium. Any
trajectory making the action stationary under the constraint of 
\begin{equation}
f\delta q+T\delta S=0\label{eq:friction-2}
\end{equation}
 satisfies Eq.~\eqref{eqn:friction} because the infinitesimal variations
$\delta S$ and $\delta q$ can be those in the direction of the time.
Change in the entropy generally should depend on how the oscillator
is moved in the time interval from $t_{{\rm init}}$ to $t_{{\rm fin}}$,
in terms of the thermodynamics. Thus, for example, the constraint
of $\delta q=0$ at $t_{{\rm fin}}$ in the variational calculation
does not always mean that of $\delta S=0$ at $t_{{\rm fin}}$. Hence, then
the constraint Eq.~\eqref{eq:friction-2} is generally nonholonomic
because it cannot be rewritten in a form that the infinitesimal
variation of a function, $U,$ of $q$ and $S$ always vanishes, i.e.,
$\delta U=0$\cite{Goldstein}.

The variation of the action Eq.~\eqref{eqn:Oaction} yields 
\begin{equation}
\int_{t_{{\rm init}}}^{t_{{\rm fin}}}\!\!\!\! dt\left\{ \left(-m\frac{d^{2}q}{dt^{2}}-kq\right)\delta q-T\delta S\right\} =0\ .\label{eqn:Oaction-1}
\end{equation}
Here, we fix both ends of the trajectory of $q$, i.e., 
\begin{equation}
\delta q(t_{{\rm init}})=\delta q(t_{{\rm fin}})=0\ .\label{eq:bq}
\end{equation}
 Substituting Eq.~\eqref{eq:friction-2} into Eq.~\eqref{eqn:Oaction-1}
gives the balance equation of momentum, 
\begin{equation}
m\frac{d^{2}q}{dt^{2}}=-kq+f\ .\label{eq:Omotiom}
\end{equation}
 The frictional force, $f$, is given by $-\xi dq/dt$ with $\xi$
being the friction coefficient, according to the linear phenomenological
law. This simple model can be regarded as a simple Kelvin-Voigt model
of the viscoelasticity. If the damping follows the Maxwell
model, the second term on the left-hand side (lhs) of Eq.~\eqref{eqn:friction}
should be replaced by the frictional force multiplied by the deformation
rate of the viscous part (dash-pot). This model is discussed in \S
\ref{sec:Viscelastic}.

Let us consider whether a straightforward way of applying the method
of undetermined multipliers to the constraint Eq.~\eqref{eqn:friction}
works well. Modifying the action so that it contains the product
of the multiplier and the lhs of Eq.~\eqref{eqn:friction}, we encounter
\begin{equation}
\int_{t_{{\rm init}}}^{t_{{\rm fin}}}dt\ \left\{ \xi\frac{d^{2}q}{dt^{2}}\delta q+\left(\frac{dT}{dS}-\frac{dT}{dt}\right)\delta S\right\} +\left[-\frac{d^{2}q}{dt^{2}}\delta q\right]_{t_{{\rm init}}}^{t_{{\rm fin}}}\!\!\!\!\!+\left[T\delta S\right]_{t_{{\rm init}}}^{t_{{\rm fin}}}=0\ ,\label{eq:holonomic0}
\end{equation}
 where we assumed that $T$ is a function of $S$, in calculating
the stationary condition of the modified action. Because the last
term on the lhs of Eq.~\eqref{eq:holonomic0} does not always vanish
under Eq.\eqref{eq:bq}, it is found that we cannot derive Eq.~\eqref{eq:Omotiom}
in this way.

\subsection{Viscous fluid\label{sec:Variational-principle-for}}

For a viscous fluid, we write $\sigma_{ij}$ and $\boldsymbol{J}_{q}$
for the viscous stress tensor and the heat flux, respectively. Let
us define the rate-of-strain tensor $e_{ij}$ as $(\partial_{j}v_{i}+\partial_{i}v_{j})/2$,
and the equation corresponding to Eq.~\eqref{eqn:friction} turns
out to be 
\begin{equation}
\rho T\ (\partial_{t}+\boldsymbol{v}\cdot\nabla)s-\sigma_{ij}e_{ij}+\nabla\cdot\boldsymbol{J}_{q}=0\ .\label{eq:Tds}
\end{equation}
 We assume the no-slip boundary condition 
\begin{equation}
\boldsymbol{v}=\boldsymbol{0}\ \ {\rm on}\ \ \partial V\ ,\label{eq:no-slip}
\end{equation}
 instead of Eq.~\eqref{eq:boundary of v}, and thus, we can assume
no dissipation at the boundary. We assume that the whole fluid is
enclosed by an adiabatic wall, i.e., that no heat flux cannot pass through the boundary. It enables us to integrate Eq.~\eqref{eq:Tds} over
$V$ to obtain 
\begin{equation}
\int_{V}\!\! d^{3}\boldsymbol{a}\ \left\{ J\left(\rho T\frac{\partial s}{\partial\tau}+\boldsymbol{f}\cdot\frac{\partial\boldsymbol{X}}{\partial\tau}\right)\right\} =0\ .\label{eq:Tds1}
\end{equation}
 Here, $\boldsymbol{f}=\nabla\cdot\sigma^{T}$ is the dissipative
force per unit volume, with the superscript $^{T}$ indicating the
transposition. In the Lagrangian description, as in Eq.~\eqref{eq:friction-2},
we thus impose the nonholonomic constraint, 
\begin{equation}
\int_{V}\!\! d^{3}\boldsymbol{a}\ \left\{ J\left(\rho T\delta s+\boldsymbol{f}\cdot\delta\boldsymbol{X}\right)\right\} =0\ ,\label{eq:Tdeltas}
\end{equation}
 instead of the adiabatic condition Eq.~\eqref{eq:adiabatic} for
the perfect fluid. Thus, the action, Eq.~\eqref{eq:L-action}, for
the perfect fluid can be rewritten into the action for the viscous
fluid as
\begin{equation}
\int_{t_{{\rm init}}}^{t_{{\rm fin}}}\!\!\!\! d\tau\!\int_{V}\!\! d^{3}\boldsymbol{a}\ \left\{ J{\cal L}(\rho,s,\boldsymbol{v})+\boldsymbol{\gamma}\cdot(\partial_{\tau}\boldsymbol{X}-\boldsymbol{v})+K(\rho J-\rho_{{\rm init}})\right\} \ ,\label{eq:SLvf}
\end{equation}
 which is a functional of $\boldsymbol{\gamma}$, $K$, $\rho$, $s$,
$\boldsymbol{v}$ and $\boldsymbol{X}$. 
The stationary condition of this action under the constraint, Eq.~\eqref{eq:Tdeltas}, can be obtained, as Eq.~\eqref{eq:Omotiom} is derived from Eqs.~\eqref{eqn:Oaction} and ~\eqref{eq:friction-2}.
 Thus, we obtain Eqs.~\eqref{eq:X}, \eqref{eq:mass}, \eqref{eqn:rho}, \eqref{eqn:vel} and 
\begin{equation}
\frac{\partial\gamma_{i}}{\partial\tau}=-\frac{\partial}{\partial a_{j}}\left\{ \rho\left(\frac{1}{2}\boldsymbol{v}^{2}-\epsilon+K\right)\right\} \frac{\partial J}{\partial(\partial X_{i}/\partial a_{j})}+Jf_{i}\ .\label{eqn:L-Navier}
\end{equation}
Substituting Eqs.~\eqref{eqn:rho} and \eqref{eqn:vel} into Eq.~\eqref{eqn:L-Navier},
we successfully obtain the balance equation of momentum for the viscous
fluid in the Lagrangian description, 
\begin{equation}
\rho\frac{\partial v_{i}}{\partial\tau}=-\frac{\partial p}{\partial X_{i}}+f_{i}\ .\label{eqn:varilagf}
\end{equation}

As in \S \ref{sub:Variational-perfect}, we can formulate the variational
principle in the Eulerian description. Let us rewrite Eq.~\eqref{eq:Tds1} as 
\begin{equation}
\int_{\boldsymbol{V}}\!\! d^{3}\boldsymbol{x}\ \left\{ \rho T\left(\partial_{t}s+\boldsymbol{v}\cdot\nabla s\right)+\boldsymbol{f}\cdot\boldsymbol{v}\right\} =0\ .\label{eq:Tds1-1}
\end{equation}
 From Eq.~\eqref{eq:fixing endpoints}, we can derive 
\begin{equation}
v_{j}=-\frac{\partial x_{j}}{\partial A_{i}}\frac{\partial A_{i}}{\partial t}\ ,\label{eqn:vkfrafra}
\end{equation}
 which can also be derived from differentiating $\boldsymbol{x}(\boldsymbol{A}(\boldsymbol{x},t),t)=\boldsymbol{x}$
with respect to $t$. By means of Eq.~\eqref{eqn:vkfrafra}, the
nonholonomic constraint Eq.~\eqref{eq:Tdeltas} can be rewritten
as 
\begin{equation}
\int_{\boldsymbol{V}}\!\! d^{3}\boldsymbol{x}\ \left\{ \rho T\delta s-\frac{\partial x_{j}}{\partial A_{i}}\left(\rho T\frac{\partial s}{\partial x_{j}}+f_{j}\right)\delta A_{i}\right\} =0\ .\label{eq:Tds2-1}
\end{equation}
 The action, Eq.~\eqref{eq:E-action}, can be rewritten as
\begin{equation}
\int_{t_{{\rm init}}}^{t_{{\rm fin}}}\!\!\!\! dt\int_{V}\!\!\! d^{3}\!\boldsymbol{x}\left\{ {\cal L}(\rho,s,\boldsymbol{v})+\beta_{i}(\partial_{t}A_{i}+\boldsymbol{v}\cdot\nabla A_{i})+K(\rho-\rho_{{\rm init}}J^{-1})\right\} \ ,\label{eq:SEvf}
\end{equation}
 which is a functional of $\boldsymbol{\beta}$, $K$, $\rho$, $s$,
$\boldsymbol{v}$ and $\boldsymbol{A}$. Calculating the stationary
condition of this action under the nonholonomic constraint, Eq.~\eqref{eq:Tds2-1},
we can obtain Eqs.~\eqref{eqn:rho}, \eqref{eq:fixing endpoints},
\eqref{eq:mascon}, \eqref{eq:v} and 
\begin{eqnarray}
\frac{\partial}{\partial t}\beta_{i} & = & -\nabla\cdot(\beta_{i}\boldsymbol{v})+\left(\rho\frac{\partial K}{\partial x_{j}}-\rho T\frac{\partial s}{\partial x_{j}}-f_{j}\right)\frac{\partial x_{j}}{\partial A_{i}}\nonumber \\
 & = & -\nabla\cdot(\beta_{i}\boldsymbol{v})+\left(-\rho\frac{1}{2}\frac{\partial\boldsymbol{v}^{2}}{\partial x_{j}}+\frac{\partial p}{\partial x_{j}}-f_{j}\right)\frac{\partial x_{j}}{\partial A_{i}}\ .\label{eq:beta}
\end{eqnarray}
 Following the same procedure as shown just below Eq.~\eqref{eq:beta-1},
we find that Eqs.~\eqref{eq:fixing endpoints}, \eqref{eq:v} and
\eqref{eq:beta} yield 
\begin{equation}
\rho\left\{ \frac{\partial}{\partial t}\boldsymbol{v}+\frac{1}{2}\nabla\boldsymbol{v}^{2}-\boldsymbol{v}\times(\nabla\times\boldsymbol{v})\right\} =-\nabla p+\nabla\cdot\sigma^{T}\ ,\label{eqn:navier}
\end{equation}
 which is equivalent to Eq.~\eqref{eqn:varilagf}. The viscous
stress tensor $\sigma_{ij}$ is a symmetric tensor because the
angular momentum is conserved, as shown in Appendix \ref{sec:Symmetry}.
If the viscous stress tensor $\sigma$ is given in the linear phenomenological
law, each of Eqs.~\eqref{eqn:varilagf} and \eqref{eqn:navier} represents
the Navier-Stokes equation.

\subsection{Viscoelastic fluid\label{sec:Viscelastic}}

We consider a viscoelastic fluid. We still write $e_{ij}$ for the
rate-of-strain tensor of the fluid, while we write $E_{ij}$ for the
strain tensor of the elastic part and ${\check{\sigma}}_{ij}$ for
the viscous stress tensor of the viscous part. The internal energy,
$\epsilon_{e}$, is a function of $\rho$, $s$ and $E_{ij}$, and
satisfies 
\begin{equation}
d\epsilon_{e}=-pd\rho^{-1}+Tds+\rho^{-1}\kappa_{ij}dE_{ij}\ ,\label{eq:kdE}
\end{equation}
 whereby $\kappa_{ij}$ is defined. We assume the rate-of-strain tensor
of the viscous part to be given by 
\begin{equation}
{\check{e}}_{ij}\equiv e_{ij}-(\partial_{t}+\boldsymbol{v}\cdot\nabla)E_{ij}\ .\label{eqn:rstvp}
\end{equation}
 We can assume that Eq.~\eqref{eq:Tds} holds if the second term
on its lhs is replaced by $-\check{\sigma}{}_{ij}\check{e}_{ij}$.
In the Eulerian description, the resultant equation is 
\begin{equation}
\int_{\boldsymbol{V}}\!\! d^{3}\boldsymbol{x}\ \left\{ \rho T\left(\partial_{t}s+\boldsymbol{v}\cdot\nabla s\right)+\left(\partial_{k}\check{\sigma}_{jk}+\check{\sigma}_{kl}\partial_{j}E_{kl}\right)v_{j}+\check{\sigma}_{ij}\partial_{t}E_{ij}\right\} =0\ .\label{eq:Tds1-1-1}
\end{equation}
 Using the same procedure in $\S$\ref{sec:Variational-principle-for},
we have the nonholonomic constraint, 
\begin{equation}
\int_{\boldsymbol{V}}\!\! d^{3}\boldsymbol{x}\ \left\{ \rho T\delta s-\frac{\partial x_{j}}{\partial A_{i}}\left(\rho T\frac{\partial s}{\partial x_{j}}+\frac{\partial\check{\sigma}_{jk}}{\partial x_{k}}+\check{\sigma}_{kl}\frac{\partial E_{kl}}{\partial x_{j}}\right)\cdot\delta A_{i}+\check{\sigma}_{ij}\delta E_{ij}\right\} =0\ .\label{eq:Tds2-1-1}
\end{equation}
 The Lagrangian density is given by 
\begin{equation}
{\cal L}_{{\rm e}}(\rho,\boldsymbol{v},s,E_{ij})\equiv\rho\left\{ \frac{1}{2}\boldsymbol{v}^{2}-\epsilon_{e}(\rho,s,E_{ij})\right\} \ .\label{eq:Le}
\end{equation}
 Replacing ${\cal L}$ in Eq.~\eqref{eq:SEvf} by ${\cal L}_{{\rm e}}$,
we obtain the action, the stationary condition of which with respect
to $\delta A_{i}$ turns out to be 
\begin{eqnarray}
\frac{\partial}{\partial t}\beta_{i} & = & -\nabla\cdot(\beta_{i}\boldsymbol{v})+\left(\rho\frac{\partial K}{\partial x_{j}}-\rho T\frac{\partial s}{\partial x_{j}}-\frac{\partial\check{\sigma}_{jk}}{\partial x_{k}}-\check{\sigma}_{kl}\frac{\partial E_{kl}}{\partial x_{j}}\right)\frac{\partial x_{j}}{\partial A_{i}}\nonumber\\
& = & -\nabla\cdot(\beta_{i}\boldsymbol{v})+\left(-\rho\frac{1}{2}\frac{\partial\boldsymbol{v}^{2}}{\partial x_{j}}+\frac{\partial p}{\partial x_{j}}-\frac{\partial\check{\sigma}_{jk}}{\partial x_{k}}\right)\frac{\partial x_{j}}{\partial A_{i}}\ ,\label{eq:beta-3}
\end{eqnarray}
 with the aid of Eq.~\eqref{eq:Tds2-1-1}, as Eq.~\eqref{eq:beta-1}
is derived. With respect to $\delta E_{ij}$, we have 
\begin{equation}
\kappa_{ij}=\check{\sigma}_{ij}\ .\label{eq:ELene}
\end{equation}
Using the same procedure in $\S$\ref{sec:Variational-principle-for},
we have the balance equation of the momentum in the Eulerian description,
which equals Eq.~\eqref{eqn:navier} with $\sigma_{ij}$ replaced
by $\check{\sigma}_{ij}$. Calculations in the Lagrangian description
can also be performed as in $\S$\ref{sec:Variational-principle-for}.

Within the linear phenomenological law, assuming $\check{\sigma}_{ij}$
to be linear with respect to $\check{e}_{ij}$ and $\kappa_{ij}$
with respect to $E_{ij}$, we can find 
\begin{equation}
(\partial_{t}+\boldsymbol{v}\cdot\nabla)\check{\sigma}_{ij}\label{eq:sigma}
\end{equation}
 to be linear with respect to $e_{ij}-\check{e}_{ij}$. Here, $\check{e}_{ij}$
is linear with respect to $\kappa_{ij}=\check{\sigma}_{ij}$. This
gives the viscoelasticity of the Maxwell type\cite{lodge,kawakatsu}.

Equation \eqref{eqn:rstvp} represents how the strain tensor is convected. Under mathematical notation, the last term on the right-hand side (rhs) of Eq. \eqref{eq:kdE}
should be $\rho^{-1}\kappa_{i}^{j}dE_{j}^{i}$, for example. If $\check{e}_{ij}$,
$e_{ij}$ and $E_{ij}$ are respectively replaced by $\check{e}_{j}^{i}$,
$e_{j}^{i}$ and $E_{j}^{i}$ in Eq. \eqref{eqn:rstvp}, the way of
convection of the contravariant tensor $E^{ij}$ involves the upper
convected time derivative. Hence, then what we have considered here
is the viscoelasticity of the upper convected Maxwell type within
the linear phenomenological law\cite{DoiOhnuki}.

\subsection{Two-component fluid}

We consider a two-component fluid. Let $\rho_{a}$ and $\rho_{b}$
be the mass densities of components $a$ and $b$, respectively. The
total mass density is given by $\rho\equiv\rho_{a}+\rho_{b},$ and
the mass fraction of component $a$ is $\psi\equiv\rho_{a}/\rho$.
We define the chemical potential as the Gibbs energy per unit mass,
and write $\mu$ for the difference given by subtracting the chemical
potential of the component $b$ from that of $a$. The internal energy
per mass is a function of $\rho_{a}$, $\rho_{b}$ and $s$, for which
we write $\epsilon_{2}(\rho_{a},\rho_{b},s)$. We have 
\begin{equation}
d\epsilon_{2}=\frac{p}{\rho^{2}}d\rho+\mu d\psi+Tds\ ,\label{eq:DE}
\end{equation}
 which can be rewritten in terms of $d\rho_{a}$, $d\rho_{b}$ and
$ds$ with the aid of 
\begin{equation}
d\rho=d\rho_{a}+d\rho_{b}\label{eq:Drho}
\end{equation}
 and 
\begin{equation}
d\psi=\frac{1-\psi}{\rho}d\rho_{a}-\frac{\psi}{\rho}d\rho_{b}\ .\label{eq:Dpsi}
\end{equation}

The components $a$ and $b$ have their respective velocities $\boldsymbol{v}_{a}$
and $\boldsymbol{v}_{b}$ and Lagrangian coordinates $\boldsymbol{A}_{a}$
and $\boldsymbol{A}_{b}$. The mean velocity of the fluid is given
by $\boldsymbol{v}\equiv\psi\boldsymbol{v}_{a}+(1-\psi)\boldsymbol{v}_{b}$.
The equations corresponding to Eqs.~\eqref{eq:invJ} and \eqref{eqn:vkfrafra}
can be respectively written as 
\begin{equation}
J_{c}^{-1}=\frac{\partial(A_{c1},A_{c2},A_{c3})}{\partial(x_{1},x_{2},x_{3})}\ ,
\end{equation}
 and 
\begin{equation}
v_{cj}=-\frac{\partial x_{cj}}{\partial A_{ci}}\frac{\partial A_{ci}}{\partial t}\ ,
\end{equation}
 where $c=a,b$. The Lagrangian density is given by 
\begin{equation}
{\cal L}_{2}(\rho_{a},\rho_{b},s,\boldsymbol{v}_{a},\boldsymbol{v}_{b})\equiv\frac{1}{2}\rho\boldsymbol{v}{}^{2}-\rho\epsilon_{2}(\rho_{a},\rho_{b},s)\ .\label{eq:LTF}
\end{equation}

In general, the dissipation is associated with the fluid viscosity,
the heat conduction and the mutual diffusion. Here, we consider a polymer
solution, where the components $a$ and $b$ are a polymer and a solvent,
respectively \cite{DoiOhnuki}. For simplicity, we neglect the
polymer elasticity and fluid viscosity, although they can be considered
as in the preceding subsections. We have 
\begin{equation}
\rho T\ \left(\partial_{t}+\boldsymbol{v}\cdot\nabla\right)s+\boldsymbol{\xi}\cdot(\boldsymbol{v}_{a}-\boldsymbol{v}_{b})+\nabla\cdot\boldsymbol{J_{q}}=0\ ,\label{eq:Tds-1}
\end{equation}
 where $\boldsymbol{\xi}$ denotes the frictional force. As we derive
Eq.~\eqref{eq:Tds2-1}, we can obtain 
\begin{equation}
\int_{\boldsymbol{V}}\!\! d^{3}\boldsymbol{x}\left\{\ \rho T\delta s-\sum_{c=a,b} \frac{\partial x_{cj}}{\partial A_{ci}}\left(\rho_{c}T\frac{\partial s}{\partial x_{j}}+\xi_{cj}\right)\delta A_{ci}\right\}=0\ ,\label{eq:TdsTF}
\end{equation}
 where $\boldsymbol{\xi}_{a}$ and $\boldsymbol{\xi}_{b}$ are introduced
to satisfy $\boldsymbol{\xi}_{a}=-\boldsymbol{\xi}_{b}=\boldsymbol{\xi}$.
As in $\S$\ref{sec:Variational-principle-for}, the action is a
functional of $\boldsymbol{\beta}_{c}$, $K_{c}$, $\rho_{c}$, $s$,
$\boldsymbol{v}_{c}$ and $\boldsymbol{A}_{c}$ with $c$ running
from $a$ to $b$, and given by 
\begin{equation}
\int_{t_{{\rm init}}}^{t_{{\rm fin}}}\!\!\!\! dt\int_{V}\!\! d^{3}\!\boldsymbol{x}\ \left[{\cal L}_{2}+\sum_{c=a,b}\left\{ \beta_{ci}(\partial_{t}A_{ci}+\boldsymbol{v}_{c}\cdot\nabla A_{ci}+K_{c}(\rho_{c}-\rho_{c\ {\rm init}}J_{c}^{-1})\right\} \right]\ ,\label{eq:SEvf-2}
\end{equation}
 where $\rho_{c\ {\rm init}}$ is the initial mass density of the component
$c$. Under the nonholonomic constraint Eq.~\eqref{eq:TdsTF},
we find the stationary condition of Eq.~\eqref{eq:SEvf-2} to be
\begin{eqnarray}
 &  & \rho_{c}-\rho_{c\ {\rm init}}J_{c}^{-1}=0\ , \\
 &  & \rho_{c}\boldsymbol{v}+\beta_{ci}\nabla A_{ci}=\boldsymbol{0}\ ,\label{eq:vA}\\
 &  & \frac{\partial}{\partial t}\beta_{ci}=-\nabla\cdot(\beta_{ci}\boldsymbol{v}_{c})+\left(\rho_{c}\frac{\partial K_{c}}{\partial x_{j}}-\rho_{c}T\frac{\partial s}{\partial x_{j}}-\xi_{cj}\right)\frac{\partial x_{cj}}{\partial A_{ci}}
\end{eqnarray}
and
\begin{equation}
K_{c}=\left(\frac{1}{2}\boldsymbol{v}^{2}-\boldsymbol{v}_{c}\cdot\boldsymbol{v}\right)+\epsilon+\frac{p}{\rho}+\mu_{c}\left(1-\frac{\rho_{c}}{\rho}\right)\ ,
\end{equation}
where $\mu_{a}$ and $\mu_{b}$ are introduced to satisfy $\mu_{a}=-\mu_{b}=\mu$.
As in $\S$\ref{sec:Variational-principle-for}, we can derive the
balance equations of momentum for the components, 
\begin{equation}
\rho_{c}\left\{ \frac{\partial}{\partial t}\boldsymbol{v}+\frac{1}{2}\nabla\boldsymbol{v}^{2}-\boldsymbol{v}_{c}\times(\nabla\times\boldsymbol{v}))\right\} =-\frac{\rho_{c}}{\rho}\nabla p+\frac{\rho_{a}\rho_{b}}{\rho}\nabla\mu_{c}+\boldsymbol{\xi}_{c}\ .\label{eqn:navier-1-3}
\end{equation}
 The sum of Eq.~\eqref{eqn:navier-1-3} over $c=a,b$ gives Eq.~\eqref{eqn:navier}
with $\nabla\cdot\sigma^{T}$ deleted. The resultant equation describes
the momentum balance of the solution as a whole. The calculations
in this subsection are in the Eulerian description; those in the Lagrangian
description can be performed as in \S \ref{sec:Variational-principle-for}.

\section{Hamiltonian formulation\label{sec:Hamiltonian-formulations-for} }

Applying the control theory \cite{Pontryagin,Schulz} to the dynamics
of the perfect fluid, we derived its Hamiltonian formulation\cite{FukagawaFujitani}.
Similarly, with slight modifications as shown below, we can derive
the Hamiltonian formulation for dissipative dynamics. In the control
theory, a subsystem controlled by another subsystem is called a plant.
Let $q$ and $u$ denote the plant state and the input to the plant,
respectively. We assume that the plant dynamics follows 
\begin{equation}
\frac{d}{dt}q(t)-F(q(t),u(t))=0\ ,\label{eq:motion of states-1}
\end{equation}
 whereby a function $F$ is defined. In addition to $q$ and $u$, we assume that the Lagrangian, $L$, also depends on a variable $S$, which will denote the entropy in the later application to dissipative dynamics. 
 In the control theory, we define an evaluation
functional 
\begin{equation}
\int_{t_{{\rm init}}}^{t_{{\rm fin}}}\!\!\!\! dt\ L(q,u,S)\ ,\label{eqn:Oaction-2-1}
\end{equation}
 and determine the optimal input $u^{*}$ so that the evaluation functional
is minimized under the constraint given by Eq.~\eqref{eq:motion of states-1}.
We also require Eq.~\eqref{eq:friction-2} and $T\equiv -{\partial L}/{\partial S}$.

We take the constraints into account using the undetermined
multiplier, $p$, which is called costate in the control theory. The
functional to be minimized, which we call action, can be defined as
\begin{equation}
\int_{t_{{\rm init}}}^{t_{{\rm fin}}}\!\!\!\! dt\ \left\{ L(q,u,S)+p\left(\frac{dq}{dt}-F(q,u)\right)\right\} \ .\label{eq:action}
\end{equation}
 Let $u^{\sharp}({q},{p},S)$ denote the input minimizing Eq.~\eqref{eq:action}
on condition that ${q},{p}$ and $S$ are fixed. This means that the
time integral of $\tilde{H}\equiv-L({q},{u},S)+pF(q,u)$ from $t_{{\rm init}}$
to $t_{{\rm fin}}$ is stationary with respect to $u$ at $u={u}^{\sharp}(q,p,S)$.
Let us define
\begin{equation}
H({q},{p},S)\equiv\tilde{H}(q,u^{\sharp}(q,p,S),p,S)\ ,\label{eq:OptHam-1}
\end{equation}
which is usually called the Hamiltonian. The preoptimized action is defined as 
\begin{equation}
\int_{t_{{\rm {init}}}}^{t_{{\rm {fin}}}}\!\!\!\! dt\ \left\{ -H(q,p,S)+p\frac{d}{dt}q\right\} \ ,\label{eq:action-2}
\end{equation}
 which is a functional of $q$, $S$ and $p$. The optimal input $u^{*}$
is given by $u^{\sharp}(q,p,S)$, when $q$, $p$ and $S$ satisfy
the stationary condition of Eq.~\eqref{eq:action-2} under Eqs.~\eqref{eq:friction-2}
and \eqref{eq:bq}. We can usually assume $\partial^2L/(\partial S\partial u)$ to vanish, and thus ${u}^{\sharp}$ is a function of only $q$ and $p$. In this case, we find 
\begin{equation}
\frac{\partial H}{\partial S}=-\frac{\partial L}{\partial S}=T\ .\label{eq:HES}
\end{equation} 
Using Eqs.~\eqref{eq:friction-2}, \eqref{eq:bq}, \eqref{eq:action-2} and \eqref{eq:HES}, we obtain a set of modified Hamilton's equations,
which are Eq. \eqref{eqn:friction}, 
\begin{equation}
\frac{d{q}}{dt}=\frac{\partial H}{\partial p}\label{eq:q-1}
\end{equation}
 and 
\begin{equation}
\frac{d{p}}{dt}=-\frac{\partial H}{\partial q}+f\ .\label{eq:p-1}
\end{equation}
 In a straightforward way from Eqs.~\eqref{eqn:friction}, \eqref{eq:q-1}
and \eqref{eq:p-1}, we have $dH/dt=0$.
 Further discussion is given in Appendix \ref{sec:Symmetry}.

Let us apply the formulation above to the viscous fluid discussed
in \S \ref{sec:Variational-principle-for}. In the Eulerian description,
the state, costate, entropy density and input are respectively given
by $\boldsymbol{A},\boldsymbol{\beta},s\ {\rm and}\ \boldsymbol{v}$.
Equation \eqref{eq:motion of states-1} is given by \eqref{eq:fixing endpoints}
in this fluid, where the function corresponding to ${\tilde{H}}$
is given by 
\begin{eqnarray}
\tilde{{\cal H}}(\boldsymbol{A},\boldsymbol{v},\boldsymbol{\beta},s) & \equiv & -\rho\left\{ \frac{1}{2}\boldsymbol{v}^{2}-\epsilon(\rho,s)\right\} +\beta_{i}\boldsymbol{v}\cdot\nabla A_{i}\ .\label{eq:hamiltonian0}
\end{eqnarray}
 Here, $\rho$ is regarded as a function of $\boldsymbol{A}$ given
by Eq.~\eqref{eq:mascon} without the undetermined multiplier $K$
introduced. Let $\boldsymbol{v}^{\sharp}(\boldsymbol{A},\boldsymbol{\beta})$
denote the input corresponding to $u^{\sharp}$, and we find the
Hamiltonian to be given by the sum of the kinetic and inertial-energy
densities, 
\begin{equation}
{\cal H}(\boldsymbol{A},\boldsymbol{\beta},s)\equiv\rho\left\{ \frac{1}{2}\boldsymbol{v}^{\sharp2}+\epsilon(\rho,s)\right\} \ ,\label{eq:opthamiltonian}
\end{equation}
 where $\boldsymbol{v}^{\sharp}$ satisfies Eq.~\eqref{eq:v}. The
modified Hamilton's equations can be found to be Eqs.~\eqref{eq:fixing endpoints}
and \eqref{eq:beta}. In deriving these equations, we note that, for
example, 
\begin{equation}
\frac{\delta}{\delta\beta_{i}}=\frac{\partial}{\partial\beta_{i}}-\frac{\partial}{\partial x_{j}}\frac{\partial}{\partial(\partial_{j}\beta_{i})}
\end{equation}
 should replace $\partial/\partial p$ in Eq.~\eqref{eq:q-1} because
the fields are involved here unlike in Eqs.~\eqref{eq:q-1} and \eqref{eq:p-1}.
In the Lagrangian description, $\boldsymbol{v}^{\sharp}$ satisfies
Eq.~\eqref{eqn:vel}, and the modified Hamilton's equations are found
to be Eqs.~\eqref{eq:X} and \eqref{eqn:varilag-1} with $\boldsymbol{v}$
replaced by $\boldsymbol{v}^{\sharp}$.

\section{Discussion\label{sec:Discussion}}

It is natural that the nondissipative equation can be derived by means of the least-action principle. We can formulate dissipative dynamics
by adding the dissipative force to the equation of motion to this
nondissipative equation. In this formulation, the variational principle
is inherent to the nondissipative dynamics. We can obtain the dissipative
force by means of the maximum dissipation principle, where the linear
phenomenological law is simply derived from the stationary condition
of the quadratic form. These two variational principles have been
utilized especially in formulating dynamics of complex fluids, as
mentioned in \S \ref{sec:Introduction}.

In the irreversible thermodynamics for a one-component viscous fluid
\cite{onsager,Degroot}, we start with the mass conservation law,
the balance equations of momentum and energy. Then, combining these
equations with the local equilibrium yields Eq.~\eqref{eq:Tds}
from which the rate of the total entropy production is written in
terms of fluxes and thermodynamic forces\cite{Youhei}. The linearity
between them is assumed to give the linear phenomenological law, determining
the dissipative force near the equilibrium.

Suppose that we do not know the balance equation of momentum but know
Eq.~\eqref{eq:Tds}, representing how the dissipation occurs. Still,
we can calculate the rate of the total entropy production from Eq.~\eqref{eq:Tds},
and then we can determine the dissipative force within the linear
phenomenological law. On the other hand, we can derive the balance
equation of momentum from Eq.~\eqref{eq:Tds} by retracing a part
of the calculations mentioned in the preceding paragraph. What we
do in this work is to formulate this derivation in a single variational
principle.

This variational principle is easy to understand in terms of physics.
The least-action principle for the perfect fluid is associated with
the entropy conservation law. Replacing this constraint on the entropy
with a constraint describing how the dissipation changes the entropy,
we can obtain our variational principle. The key point inherent to
the dissipative dynamics is to appreciate that this is the nonholonomic
constraint to which the method of undetermined multipliers cannot
be applied.

Applications of this variational principle to a viscous fluid, a viscoelastic
fluid and a two-component fluid are shown in \S \ref{sub:Dissipative}.
They are straightforward generalizations of the simple case discussed
in \S \ref{simple}. We believe that this principle can be easily
applied to the dynamics of other complex fluids, such as liquid
crystals and electromagnetic fluids\cite{liusensei,liu2}.

\section*{Acknowledgements}

H. F. was supported in part by a KLL Research Grant for
Ph.D. Program and a Grant-in-Aid for the GCOE Program for the Center for
Education and Research of Symbiotic, Safe and Secure System Design
from MEXT, Japan. The authors thank C. Liu, T. Kambe, and T.
Yamaguchi for valuable discussions. H. F. also thanks S. Goto for valuable
comments.

\appendix

\section{Symmetry\label{sec:Symmetry}}

We show that the Hamiltonian version of Noether's theorem\cite{Goldstein,fukaya}
still works for the model discussed in \S \ref{sec:Hamiltonian-formulations-for}.
Let us consider a canonical transformation, mapping $(q_{0},p_{0},S_{0})$
to $(q_{\alpha},p_{\alpha},S_{\alpha})$, which makes the modified
Hamilton's equations, Eqs.~\eqref{eqn:friction}, \eqref{eq:q-1} and
\eqref{eq:p-1}, invariant. When the transformation with respect to $\alpha$
is infinitesimal, the generator $G(q_{0},p_{\alpha},S_{0},t)$ can be defined.
Then, because of the canonical transformation, we have 
\begin{equation}
\int_{t_{{\rm init}}}^{t_{{\rm fin}}}\!\!\!\! dt\ \left\{ p_{0}\dot{q}_{0}-H_{0}(q_{0},p_{0},S_{0})\right\} =\int_{t_{{\rm init}}}^{t_{{\rm fin}}}\!\!\!\! dt\ \left\{ p_{\alpha}\dot{q}_{\alpha}-H_{\alpha}(q_{\alpha},p_{\alpha},S_{\alpha})+\frac{d}{dt}\left(q_{0}p_{\alpha}+\alpha G\right)\right\} \ ,\label{eq:HH-1}
\end{equation}
 where $(q_{0},p_{0},S_{0})$ and $(q_{\alpha},p_{\alpha},S_{\alpha})$
respectively satisfy the modified Hamilton's equations. With the
aid of Eqs.~\eqref{eqn:friction} and \eqref{eq:bq}, Eq.~\eqref{eq:HH-1}
yields 
\begin{eqnarray}
\delta{q} & \equiv & {q}_{\alpha}-{q}_{0}=\alpha\frac{\partial G}{\partial p_0}\ ,\label{eq:Gq}\\
\delta{p} & \equiv & {p}_{\alpha}-{p}_{0}=\alpha\left(-\frac{\partial G}{\partial q_0}+\frac{f}{T}\frac{\partial G}{\partial S_0}\right)\label{eq:Gp}
\end{eqnarray}
 and
\begin{equation}
H_{\alpha}(q_{\alpha},p_{\alpha},S_{\alpha})-H_{0}(q_{0},p_{0},S_{0})=\alpha\frac{\partial G}{\partial t}\ ,\label{eq:HH}
\end{equation}
where we simply wrote $G$ for $G(q_0,p_0,S_0,t)$. From Eqs.~\eqref{eq:friction-2}, \eqref{eq:HES}--\eqref{eq:p-1},
\eqref{eq:Gq} and \eqref{eq:Gp}, we have 
\begin{eqnarray}
H_{\alpha}(q_{\alpha},p_{\alpha},S_{\alpha}) & = & H_{\alpha}(q_{0},p_{0},S_{0})+\frac{\partial H}{\partial q}\delta q+\frac{\partial H}{\partial p}\delta p+\frac{\partial H}{\partial S}\delta S\label{eq:dH}\\
 & = & H_{\alpha}(q_{0},p_{0},S_{0})-\alpha\left(\frac{dp}{dt}\frac{\partial G}{\partial p}+\frac{dq}{dt}\frac{\partial G}{\partial q}+\frac{dS}{dt}\frac{\partial G}{\partial S}\right)\ .\label{eq:dH1}
\end{eqnarray}
 Substituting Eq.~\eqref{eq:dH1} into Eq.~\eqref{eq:HH} yields
\begin{equation}
H_{\alpha}(q_{0},p_{0},S_{0})-H_{0}(q_{0},p_{0},S_{0})=\alpha\frac{dG}{dt}\ .\label{eq:identityH}
\end{equation}
Thus, the condition $dG/dt=0$ is equivalent to the invariance of $H$, i.e., $H_{\alpha}=H_{0}$, under the canonical transformation. If time-invariant, $H$ is the generator of the
infinitesimal canonical transformation with respect to the time.

As discussed in $\S$\ref{sec:Hamiltonian-formulations-for}, the
Hamiltonian density of a fluid is given by ${\cal H}(\boldsymbol{A},\boldsymbol{\beta},s)$
of Eq.~\eqref{eq:opthamiltonian}, where $\boldsymbol{A}$, $\boldsymbol{\beta}$ and ${s}$
are respectively the state, costate and entropy density. As Eq.~\eqref{eq:identityH},
we can derive
\begin{equation}
\int_{V}\!\! d^{3}\!\boldsymbol{x}\ \left\{ {\cal H}_{\alpha}(\boldsymbol{A}_{0},\boldsymbol{\beta}_{0},s_{0})-{\cal H}_{0}(\boldsymbol{A}_{0},\boldsymbol{\beta}_{0},s_{0})\right\} =\int_{V}\!\! d^{3}\!\boldsymbol{x}\ \alpha\frac{d{\cal G}}{dt}\ .\label{eq:identityH-1}
\end{equation}
 Thus, if ${\cal H}$ is invariant under the transformation
generated by ${\cal G}$, the integral of ${\cal G}$ over $V$ is conserved.
This means that a vector field $\boldsymbol{J}$ can be defined so that 
\begin{equation}
\frac{\partial{\cal G}}{\partial t}+\nabla\cdot\boldsymbol{J}=0\label{eq:Gt}
\end{equation}
 is satisfied and that the surface integration of $\boldsymbol{J}\cdot\boldsymbol{n}$
over $\partial V$ vanishes. Its converse also holds. The spatial
transformation is generated by the momentum, $\rho\boldsymbol{v}^{\sharp}=-\beta_{i}\nabla A_{i}$. The total momentum of the fluid is conserved when there are no external forces acting on the boundary ${\partial V}$. It is thus natural that we can rewrite Eq.~\eqref{eqn:navier}
in the form of Eq.~\eqref{eq:Gt}, i.e., 
\begin{equation}
\frac{\partial}{\partial t}(\rho v_{i}^{\sharp})+\partial_{j}(\rho v_{i}^{\sharp}v{}_{j}^{\sharp}-T_{ij})=0\ ,\label{eq:moment}
\end{equation}
 where $T_{ij}\equiv p\delta_{ij}-\sigma_{ij}$ is a stress tensor.
The spatial rotation is generated by the angular momentum, $\boldsymbol{l}\equiv\rho\boldsymbol{x}\times\boldsymbol{v}^{\sharp}=-\boldsymbol{x}\times\beta_{i}\nabla A_{i}$.
The total angular momentum of the fluid is conserved when there is no external
torque acting on the boundary ${\partial V}$. The time evolution of $\boldsymbol{l}$ turns out
to be 
\begin{equation}
\frac{\partial l_{i}}{\partial t}+\frac{\partial}{\partial x_{l}}\left(l_{i}v_{l}^{\sharp}-\epsilon_{ijk}x_{j}T_{kl}\right)-\epsilon_{ijk}T_{kj}=0\ ,\label{eq:moment-1-2-1}
\end{equation}
 with the aid of Eq.~\eqref{eq:moment}, where the Levi-Civita symbol
$\epsilon_{ijk}$ is anti-symmetric on each pair of indices. The last
term on the lhs above should vanish from the discussion just around
Eq.~\eqref{eq:Gt}, which leads to $\sigma_{ij}=\sigma_{ji}$. 

\end{document}